\documentclass{cimento} 
\usepackage{AMSmath}
\usepackage{graphicx}
\usepackage{bm}

\title{Detecting gravitational waves with a heralded-photon Quantum Eraser}
\author{
F.~Tamburini\from{ins:1}\ETC
C.~J.~de~Matos\from{ins:2}
J.~M.~Perdigues~Armengol\from{ins:3}
\atque
C.~N.~Colacino\from{ins:4}}
\instlist{
 \inst{ins:1} Dept. of Astronomy, University of Padova - vicolo dell'Osservatorio 3, IT-35122 Padova, Italy
 \inst{ins:2} ESA-HQ, European Space Agency - 8-10 rue Mario Nikis, 75015 Paris, France
 \inst{ins:3} ESA-ESTEC, European Space Agency - Keplerlaan 1, 2200 AG Noordwijk ZH, the Netherlands
 \inst{ins:4} Dept. of Physics, University of Pisa, Largo B. Pontecorvo 3, I-56127, Pisa, Italy
}

\PACSes{
\PACSit{03.65.Ud}{Entanglement and quantum nonlocality}
\PACSit{04.80.Nn}{Gravitational wave detectors and experiments}
\PACSit{03.67.-a}{Quantum information}
}

\begin{document}

\maketitle

\begin{abstract}
We propose the use of heralded photons to detect Gravitational Waves (GWs). Heralded photons are those photons that, produced during a parametric downconversion process, are ``labelled'' by the detection and counting of coincidences of their correlated or entangled twins and therefore can be discriminated from the background noise, independently of the type of correlation/entanglement used in the setup.
Without losing any generality, we illustrate our proposal with a \textit{``gedankenexperiment''}, in which the presence of a gravitational wave causes a relative rotation of the reference frames associated to the double-slit and the test polarizer, respectively, of a Walborn's quantum eraser \cite{wal02}. In this thought experiment, the GW is revealed by the detection of heralded photons in the dark fringes of the recovered interference pattern by the quantum eraser.
Other types of entanglement, such as momentum-space or energy-time, could be used to obtain heralded photons to be used in the future with high-frequency GW interferometric detectors when enough bright sources of correlated photons will be available.
\end{abstract}

\section{Introduction}
The detection of Gravitational Waves represents one of the most fascinating and challenging targets of modern physics \cite{blair,saulson}.
GWs are wave-like solutions of Einstein's General Relativity in the weak-field limit, i.e. when the spacetime metric can be written as the sum of a background, time-independent part, $g_{\mu \nu }^{(0)}$ and a 
perturbation $h_{\mu \nu }$, with $|h|<<1$ \cite{sw72}. To detect GWs, experimenters have to take into account mainly two constraints in order to build the best-optimized detector. 
The first is the weak coupling of gravitational radiation with matter and the second is the estimation of waveform, frequency and amplitude of GWs emitted by astrophysical sources. Different types of sources are thought to generate different types of gravitational radiation, e.g., short bursts with relatively high frequency waves can be generated by catastrophic collisions of compact objects like supernovae events, neutron star oscillations or other types of oscillating mass distributions (pulsars, binary systems, etc.). Long wavelength spacetime ripples, instead, are thought to have been created mainly during the early universe epoch or generated by cosmological sources such as galactic black holes \cite{blair}.

Current and next-generation ground-based experiments \cite{weber,ligo1,ligo2,ligo3,geo1,geo2,virgo1,virgo2,virgo3} intend to measure microscopic deformations of the order of the GW amplitude $h$ in cryogenic rods or in the paths of laser interferometers, that barely correspond to a fraction of the classical size of a proton.
Although Hulse and Taylor have provided an indirect historical proof of their existence with their radio-astronomical observation of the famous double pulsar PSR1913+16 \cite{hulsetaylor,weistay}, no successful detection claim has been made so far. The sensitivity of current and future ground-based interferometric detectors is limited mostly by three major noise sources: the low-frequency limit, $\nu\sim10$Hz is dictated by seismic noise, ie the noise due to the Earth's gravitational mass. The high-frequency limit, $\nu\sim10$kHz is dictated by the shot-noise, ie the intrinsically quantum noise which comes from the Poissonian fluctuations in the number of photons impinging onto the mirrors. At intermediate frequencies, it is the thermal noise, ie the noise of the mirror suspensions and coatings , which dominates. In order to overcome the seismic barrier, the LASER Interferometer Space Antenna (LISA) mission has been envisaged \cite{lisa1}: LISA is made up by three free-falling satellites which will orbit around the sun and will probe the frequency region $30$nHz-$0.1$Hz. The mission will start
in 2020 \cite{lisa2}.

Thermal noise will be the most important noise term for future projects, such as the Einstein Telescope (ET) \cite{et}. New suspensions and new coating materials are currently under investigation.

Quantum optics techniques have been already proposed to improve the performances of GW detectors, at high-frequencies, e.g., by using squeezed light states \cite{cav81,vahl08} to reduce the shot noise of the measurement and pushing the sensitivity over the barrier dictated by the quantum limit \cite{unr83}.
Even more exotic ideas such as quantum nondemolition measurements on the mirror test mass position \cite{kim02}, the use of entangled photon states \cite{tam00,tam08} and even the use of atom 
interferometers \cite{kasevich,colacino} have been proposed to overcome the shot noise problem. In this view, heralded photons states could actually represent a realistic improvement for the detection of GWs.

Heralded photons are produced through a parametric downconversion (PDC) process, which generates pairs of correlated photons (or more complicated sets of $N$ quanta) of which entangled states are a particular case.
In the simplest case, when $N=2$, the detection of one member of each of these pairs can be used to herald the presence of the other twin photon making the distinguishability of a photon click from the detector's background noise a reality \cite{hon86,lvo01,pit05}.

To illustrate our proposal (Fig. 1) let's take as an example a simple thought experiment where a GW introduces a relative rotation on two reference frames, one harbouring the double-slit ($R_s$) and the other the test polarizer P  ($R_p$),  building an ideal Walborn's quantum eraser \cite{wal02}.
In this experiment, entangled pairs of photons are produced from spontaneous PDC in a Beta-Barium Borate (BBO) crystal illuminated by an Argon ion pump laser. The entangled photons in each entangled pair are produced in mutually orthogonal linear polarization states, $\mid o>$ and $\mid e>$. 
The first photon (photon $s$) follows the path $s$ and goes through the double slit to create its detection event in $D_s$. The other photon (photon $p$), instead, travels through the path $p$ directly to the detector $D_p$. Finally, a coincidence counter records the correlation of events created by the detection of both the two photons of the entangled pair. 

Two Quarter-Wave Plates (QWP) are placed in front of each the two slits and are oriented so that their fast axes are mutually orthogonal. The two QWPs change the linear polarization of the incoming light beams into two differently circular polarization states, with the result of marking the paths of each interfering photon destroying the interference pattern. For this reason this experiment is also called a ``which-way experiment'', meaning that the experimenter can know after a polarization measurement the path followed by each photon that crosses the double slit. 
A ``quantum eraser'' can cancel this which-path information and restore the interference pattern through an appropriate measurement of the polarization of the other entangled photon (photon $p$).

In his ``which-way experiment'' Walborn observed that some residual interference is still present, even after the installation of the QWPs before the slits, attributing this residual interference to a small error in the alignment of the QWPs or to some residual imperfections present in the optical setup. In fact the detection of those spurious photons could have been distinguished with high confidence from the detectors' dark counts thanks to the known properties of heralded photons.
This means that, by taking into account the presence of the existing imperfections, one could either measure the precision of the QWPs alignment or that of the test polarizer P or, even, to characterize the optical quality of the hardware present in the setup.

In the ideal case, after having taken into account all the possible instrumental and environmental effects in the setup, the presence of coincidences generated by heralded photons in the dark fringes of the interference pattern restored through a quantum erasure could in principle signal the presence of GWs. 
GWs, in fact, will introduce a relative rotation $\Delta \beta$ between the two reference frames Rs and Rp proportional to the GW amplitude $h$ with the result of misaligning the quantum eraser setup and that would partially prevent the complete restoration (or destruction) of the interference pattern revealing the presence of GWs (\cite{tam08}).
Additional spurious events that might be caused by the decoherence of entanglement due to the presence of GWs can be neglected here, being the order $h^2 \ll h$ \cite{kok03,tam08}.

Of course there are some caveats that make this proposal more at a level of \textit{gedankenexperiment} than of a realistic experiment to detect GWs through a simple quantum eraser.
For example, in the classical Walborn's setup, the interference pattern is drawn by the counting of coincidences of a moving detector that should be replaced by an array of photon counters, $D_s$, to achieve a reasonably fast reading. In fact, the original setups of those experiments described by Walborn \textit{et al. }\cite{wal02}, would take a relatively long time to acquire the photons; a time too long to detect a gravitational wave burst from a supernova or to monitor the latest stages of a binary inspiral. In addition, at this stage of \textit{gedankenexperiment}, it seems unrealistic to study problems that detectors in advanced stage of design should face, like the environmental, instrumental and seismic noises.
We only point out that the order of magnitude of the ellipsometric measurement required for the detection of GWs through a measurement of the polarization states of light, even very small \cite{cru00,tam00,pra02,kok03,tam08}, can be anyway achieved with the present technology. We cite the example of an existing experiment, \textit{Polarizzazione del Vuoto con LASer} (PVLAS) designed to detect the small ellipticity ($\sim 5 \times 10^{-11}$) acquired by a linearly polarized light due to vacuum's magnetic birefringence in the presence of strong magnetic fields \cite{pvlas}.

Being based on a photon-coincidence counting procedure a heralded-photon GW detector will not be limited by shot noise, becoming a good candidate for High or Ultra-High frequency GW detectors \cite{tam00,tam08}.
The use of heralded photons, obtained from entangled photon states and other particular exotic states of entanglement, permits to overcome the shot noise problem through a dependence of $1/N$ instead of $1/\sqrt{N}$, where $N$ is the number of the quanta in each heralded (entangled) photon state.
Thus the measurement precision is limited by the number $N$ of photons used. The classical measurement setups that use each of the photons independently has a phase uncertainty $\sim 1/\sqrt{N}$, obtaining the well-known standard quantum limit. 
However, in quantum metrology, it is possible to achieve a precision ideally limited only by Heisenberg's uncertainty principle, with the result of a dramatical improvement to the $1/N$-behavior by using particular quantum entangled states such as N00N and GHZ states. 
The N00N state distribution is a particular quantum-mechanical many-body (path or polarization) entangled state proposed for super-resolution techniques such as precision phase measurements in optical interferometers and Schr\"odinger cat states. The generic N00N state is made by $N$ particles casted in a singlet of $N$ and $0$ particles,
$|\psi_N\rangle = |N\rangle_a |0\rangle_b + e^{iN\theta}|0\rangle_a|N\rangle_b$, whose
expectation value is $A=\pm 1$ and phase $\theta = [0, \pi/N]$, with an error in the phase determination $\Delta \theta=1/N$, much below the quantum limit $\sqrt{N}$.
Closely related to the N00N states are the maximally entangled Greenberger--Horne--Zeilinger (GHZ) states where any single member do not possess single particle properties but only the properties of the whole quantum state \cite{mau01,mit04,mit05,san89,ghz90}. 
Recent studies, also corroborated by experimental evidence, demonstrated the actual feasibility and difficulties of building and measuring these multi-entangled photon states in more efficient ways with the result of improving their production rate \cite{qi09,lu07,wie08}.

The adoption of heralded photons obtained from entangled pairs, N00N or GHZ states proposed by us can improve or even compete with another approach based on nonclassical properties of the electromagnetic field known as \textit{Ôsqueezed lightÕ} technique, that makes possible, with some analogies, the breaking of the standard quantum limit but, in this case, at the expense of increasing the noise in another degree of freedom \cite{slu85,wu87,ger05,bac04,sha09}.
Quadrature-phase squeezed states of light have been proposed - and tested - to improve the performances of large-scale interferometric GW as in the LIGO and VIRGO projects, with a $S/N$ ratio improvement up to the $44\%$ with respect to the shot noise \cite{god08}. 

While the main difficulty of our proposal is the generation of high-$N$ quantum states, squeezed light presents instead another side effect dictated by the limit of the light squeezing recently discussed especially for spin states, as mentioned in Refs. \cite{com05,pry09}.
It can also be argued that, while strong sources increase the probability of detecting events, they unavoidably may concur to generate spurious events also with the use of squeezed light. On the other side, we see that one can obtain similar results by using weaker sources of heralded photons, but with the key advantage of having the secure validation of a real event from random detections by controlling each of the photon states, and with the additional crucial advantage of beating the shot noise to the $1/N$ limit. Brighter sources will further improve the capabilities of this setup.

\section{Toy-model: $N=2$ and GW detection with a Wallborn Quantum eraser}

Here we mainly discuss, for the sake of simplicity and without losing any generality, the use of heralded photons obtained from maximally polarized photon entangled pairs, i.e. when $N=2$, which represents the simplest example of N00N state. Of course, GHZ states are obtained, by definition, when $N>2$.
We also prefer to focus our attention in polarization entangled states that may offer greater advantages than those given by light-squeezed polarization states  \cite{com05,pry09}.
The use of heralded photon states with the existing laser interferometry-based GW detectors, instead, should be obtained by exploiting energy-time or momentum-space correlated photon states. 

In any case, the main core of our proposal is to show that heralded photons can improve the sensitivity of the existing GW detectors, as it is also the case when squeezed light techniques are used, but offering the new crucial advantage of distinguishing a real signal from the instrumental noise even at the single photon level through the properties of quantum heralded photons.
Moreover, also in the simplest state with $N=2$ the experimenter can overcome the shot-noise problem, improving the detection of a factor $1/\sqrt2$ with respect to the quantum limit.

The source consists on a \textit{Type-II} spontaneous parametric downconversion process occurring in a non-linear crystal that creates pairs of maximally entangled photons in the singlet state:
\begin{equation}
\mid \Psi >=\frac{1}{\sqrt{2}}(\mid o>_s \mid e>_p + e^{i
\phi}\mid e>_s\mid o>_p)
\label{e1}
\end{equation}
where $o$ and $e$ are the ordinary and extraordinary polarizations, respectively, $\phi$ the relative phase shift induced by the birefringence of the crystal. 
If $\phi = 0$, then the source generates entangled photons in the Bell state
$\mid \Psi^+ >$, while the Bell state $\mid \Psi^- >$ is associated to $\phi = \pi$.

In Walborn's interferometer (Fig. 1), the probability $P_1$ of detecting heralded photon coincidences between the detectors located behind the polarizer P and the array of detectors monitoring the screen behind the double slit S is \cite{wal02}

\begin{equation}
P_1
\propto\frac{1}{2}+\Big[\frac{1}{2}-\sin^2(\theta+\alpha)
\cos^2\frac{\phi}{2}-\sin^2(\theta-\alpha)\sin^2\frac{\phi}{2}\Big]\sin
\delta\label{e2}
\end{equation}
where $\delta$ is the phase difference accumulated by the quanta between the two paths from slit 1 and from slit 2 to the position where the photon is detected after crossing the two slits. The angle $\theta$ is the smallest angle formed by the fast (slow) axis of the QWPs and the photon polarization axis $o$, while $\alpha$ is the inclination of the polarizer with respect to the photon polarization axis $o$ in the path $p$.

The basic setup of Walborn's interferometer can be modified either by removing the QWPs, or the polarizer P (or both). 
Each of these modifications will unavoidably affect the probability $P_1$ of detecting coincidences with heralded photons.
\begin{enumerate}
    \item When no QWPs are present in the path of beam $s$, no polarizer is present in the path of beam
    $p$ and $\theta=\alpha=0$, then the overlapping of the two photon paths,  $s_1$ and $s_2$, 
    generates the usual double-slit interference pattern and $P_1$ becomes
     \begin{equation}
    P_1\propto\frac{1}{2}(1+\sin{\delta})\label{e3}
    \end{equation}
    
    \item If the two QWPs are present along the path of the beam $s$, the
    polarizer P is not inserted and $\theta=\pm \frac{\pi}{4}, \alpha=0$, then
    \begin{equation}
    P_1\propto\frac{1}{2}(1+\sin{\delta} \hspace {0.1 cm} \cos{2
    \theta})=\frac{1}{2}\label{e4}
    \end{equation}
The availability of the which-path information here imposed by the two QWPs unavoidably marks the paths $s_1$ and $s_2$, leading to the destruction of the interference pattern.
With this configuration an observer has indeed the possibility (if he wishes) to gain information about the which-path information without perturbing the $s$ photon until its detection.  This is possible by measuring the polarization of the photon states associated to the paths $s$ and $p$ essentially in two ways:
    
    \textbf{a) }$p$ is measured \emph{before} $s$, 
    
    \textbf{b) }$p$ is measured \emph{after} $s$ (a.k.a. \textit{delayed erasure}).
    
    \item Finally, when both the QWPs and the Polarizer P are present and also
    $\theta=\pm\frac{\pi}{4}, \alpha=\pm\frac{\pi}{4}$, so that the polarizer will force light in a combination of $o$ and $e$ polarizations. The which-path information is erased and the interference pattern restored, according to eq. (\ref{e2}).
\end{enumerate}

In the following, we will use the situations presented in points 2 and 3 to illustrate our proposal.
As already said, the core of this \textit{gedankenexperiment} is to illustrate how to measure with heralded-photon quantum states a classical effect caused by the presence of a plane GW, namely the relative rotation $\Delta \beta \sim h$ induced between the two local reference frames $R_p$ and $R_s$. Following  \cite{tam08}, the effect of the GW corresponds to the rotation of the local polarization eigenvector basis along which the undefined polarization states of each single photon in an entangled pair are measured .

Let us consider for the sake of simplicity and, without losing any generality, the effect of a monochromatic plane GW, with wavelength $\lambda_{GW}$, propagating along the negative $x^3$ direction of the instrument's reference frame. We also assume that the coordinates are harmonic.

To obtain an omni-directional detector, we simply propose to use a different Wallborn interferometer for each of the spatial directions and associate to each of the three spatial axes their paths $p$ and then compare the results independently obtained. Cross-correlation of three independent experiments can ensnare the signature of a gravitational wave.

The geometry is described by the perturbed metric
$g_{\mu \nu}=\eta_{\mu \nu}+h_{\mu \nu}$, where $h_{\mu \nu}$ is the perturbation caused by
the GW to the Minkowsky flat metric $\eta_{\mu \nu}$. 

In our example, to detect GWs propagating along the $x^3$ axis we align the path $p$ with the $x^1$ axis and the metric tensor is
\begin{equation}
g_{\mu \nu}=\eta_{\mu \nu}+h_{\mu \nu}=\left(
\begin{array}{cccc}
1 & 0 & 0 & 0 \\
0 & -1 + h_+ \cos \Theta  & h_\times \cos \Theta  & 0 \\
0 & h_\times \cos \Theta  & -1- h_+ \cos \Theta  & 0 \\
0 & 0 & 0 & 0
\end{array}
\right)
\end{equation}
where $h_+$ and $h_\times$ represent the amplitudes of the two polarization states of the GW.
The phase of the wave at position $x^{\mu }$ is given by $\Theta =k_{\mu}\cdot x^{\mu}$, where $k^{\mu }=(\frac{2\pi }{\lambda _{g}},0,0,-\frac{2\pi}{\lambda _{g}})$ represents the wave-vector of the GW.
Let us align the two paths $|p\rangle$ and $|s\rangle$ along the $x^1$ and $x^2$ axes, respectively. The two beams are forced onto the plane $(x^1,x^2)$ orthogonal to the propagation direction of the GW and we assume that the initial polarization vectors of the two light beams are $\Pi_s= (x^0,x^1,x^2,x^3)=(1,0,1,0)$ and $\Pi_p=(1,0,0,1)$, respectively. The rotation of the polarization vector is given by the appearance of the additional components $\Pi^2$ and $\Pi^3$ measured by the detectors at the end of the two paths $|p\rangle$ and $|s\rangle$, respectively. The two paths of the beams can be parameterized in term of the affine parameter $\sigma$, which is the distance traveled by each individual photon in each different arm. The position four-vectors of paths $s$ and $p$ are $x_s=(\sigma, 0, \sigma, 0)$ and $x_p=(\sigma,\sigma,0,0)$, respectively.

The total change in polarization that can be measured by the detector along the path 
$|p\rangle$ is given by
\begin{equation}
\Delta \Pi^2_p =\int^{l_p}_0 \frac{d \Pi}{d\sigma}d\sigma= - \int^{l_p}_0 
\frac{\pi h_\times}{\lambda_{GW}} \sin  \left(\frac{2 \pi \sigma}{\lambda_{GW}} \right)  \; d\sigma
= - \frac12 h_\times \left[1- \cos \left(\frac{2 \pi l_p}{\lambda_{GW}} \right) \right]
\end{equation}
and similarly for the path $|s\rangle$ we find that $\Delta \Pi^3_s = - \frac12 h_\times \left[1- \cos \left(\frac{2 \pi l_s}{\lambda_{GW}} \right) \right]$.
As sketched in point 3, the quantum eraser can restore the interference pattern by rotating the polarizers restoring the dark fringes. When a GW is interacting with the photons in the paths $l_s$ and $l_p$, because of the factor $\Delta \beta$, the restored interference pattern will be perturbed by the GW. From a practical point of view the experimenter could detect the effect of a GW by observing the presence of heralded photons coincidences inside the dark fringe. We still point out that the properties of heralded photons will allow the experimenter to tell a real photon apart from the noise.

For small angles $\Delta \beta_s \simeq \Delta \Pi^3_s$ and $\Delta \beta_p \simeq 
\Delta \Pi^2_p$. From eq. (\ref{e2}) we obtain the probabilities associated to the Bell states $\Psi_+$ and $\Psi_-$ that, after having interacted with the GW, become
\begin{equation}
P_{\Psi_\pm}=\frac12 - \left\{ \frac12 - \cos^2\left[\frac\pi4 + \Delta \beta_s \pm (\alpha +\Delta \beta_p) \right] \right\} \sin \delta
\end{equation}
which at the first order in $\alpha$ for the $\Psi_{+}$ state, to which we will mainly refer in the paper for the sake of simplicity, becomes
\begin{equation}
P_{\Psi_+}=\frac12 - (\alpha + \Delta \beta) \sin \delta
\end{equation}
where
\begin{equation}
\Delta\beta = \Delta\beta_s+\Delta\beta_p = -\frac12 h_\times \left[ 2 - \cos \left(\frac{2 \pi l_p}{\lambda_{GW}}\right) - \cos \left(\frac{2 \pi l_s}{\lambda_{GW}}\right) \right]
\label{eqdbeta}
\end{equation}

In the \textit{delayed erasure} setup, we fix the distance between the entangled source and the slit much smaller than the path $p$, $l_s \ll l_p$, i.e., (or equivalently, in the \textit{non-delayed erasure} configuration $l_p \ll l_s$),  the absolute value of the deviation induced by a plane GW, propagating in a direction orthogonal to the photon trajectory can be approximated by \cite{cru00,tam00}
\begin{equation}
\Delta \beta \simeq h_\times \left[ 1- \left(\frac{\pi l_p}{\lambda_{GW}}\right) - \left(\frac{\pi l_s}{\lambda_{GW}} \right) \right] 
\label{e5}
\end{equation} 

Let us consider a GW detector for the high frequency GW spectrum in the kHz - GHz region. We choose $l_p=1000m$ and $l_s=31m$. 
We have analysed the response of this detector, namely the angular rotation between the two reference frames, $\Delta \beta$ for one single arm as a function of a GW having $h_\times=10^{-18}$ in the spectrum of wavelengths $300m <\lambda_{GW}<300km$. 
The results are plotted in Fig. 2. In all our simulations the three last peaks observed before the asymptotic regime remain quite stable for any value of the ratio $l_p/l_s \gg 1$.

Following \cite{tam08}, this setup can be optimized to work at the so-called \textit{long wavelength limit}, i.e. for GWs with $\lambda_{GW}\gg l$. After the broader peak found at 10MHz, the response stabilizes to an asymptotic regime with $\Delta \beta \sim 10^{-4} h_\times$, similar to the response obtained from the ground-based GW detectors without the limitation of the shot noise, being the detection at the single photon regime.
Metric perturbations add monotonically giving rise to an asymptotically linear regime, as can be easily seen from the first-order expansion in eq. \ref{e5}, and reported in the main body of Fig. 2. 

In the UHF spectrum, corresponding to the \textit{short wavelength limit} of the detector, instead, simulations show that the plot starts being populated by a forest of narrower and narrower peaks due to the finiteness of the ratio $\l_p/l_s$ (see the inset of Fig. 2).
Plane GWs with wavelength $\lambda_{GW} \leq l_p$ interact with the detector, but their effects do not add monotonically because the periodic metric perturbations caused by the plane wave periodically rotate and then restore the alignments between $R_p$ and $R_s$. As shown in the inset of Fig. 2, the angle $\Delta \beta$ presents several oscillations.

In this case, the optimal sensitivity will be achieved for all plane GWs having wavelength for which the cosines go to zero, which happens if
\begin{equation}
\lambda_{GW}=\frac{4l_{p,s}}{2k_{p,s}+1}
\end{equation}
where $k_{p,s}$ are positive integers corresponding for the paths $l_s$ and $l_p$, respectively.
The maximum effect is obtained for $k=1$, when the arm $P$ of the detector is $\frac 34$ times the GW wavelength. 

In the particular case, when the paths $l_p$ and $l_s$ are both multiple of the GW wavelength,  e.g. $l_p = n \lambda_{GW}$, ($n$ is a positive integer), the rotation of the polarization vector is fully restored along the path and $\Delta \beta = 0$, and thus the GW cannot be detected as already discussed in Ref. \cite{kok03}. 
To avoid this problem present in the short wavelength regime, experimenters should choose a setup where $l_p/l_s$ is irrational. In fact, from eq. \ref{eqdbeta}, we find $\Delta \beta=0$ iff $ 2 - \cos \left(\frac{2 \pi l_p}{\lambda_{GW}}\right) - \cos \left(\frac{2 \pi l_s}{\lambda_{GW}}\right)=0$, that happens when both the cosines assume the positive unity value, i.e. when 
\begin{eqnarray}
\frac{2 \pi l_s}{\lambda_{GW}}=\frac{2k_s+1}{2}\pi \label{condcos}
\\
\frac{2 \pi l_p}{\lambda_{GW}} = \frac{2k_p+1}{2}\pi \nonumber
\end{eqnarray}
where $k_s$ and $k_p \in N$, i.e. are integer positive numbers. If the ratio $l_p/l_s$ is a positive integer, by assuming $l_p>l_s$ and, therefore, $k_s\geq k_p$, we obtain from the direct sum of equations \ref{condcos} that
\begin{equation}
\frac{l_s}{\lambda_{GW}}\left(1+ \frac{l_p}{l_s} \right)=2(k_p+k_s+1) \geq 2 \; \;  (\in N)
\end{equation}
which admits $k_s-1$ solutions when $l_p/l_s$ is a positive integer.

By varying the length of the arm $P$, the observer can tune the detector at a specific wavelength $\lambda_{GW}$ and observe also all the harmonics obtained for all values of $k>1$, drawing with precision the entire GW spectrum with the three main peaks before the long wavelength limit.
The photon path in the arm $P$ behaves like a string tuned to a specific GW wavelength $\lambda_{GW}$ and to its superior harmonics.
The analogy of this detector with the baroque string instrument with sympathetic strings, \textit{viola d' amore}, is quite straightforward.

\begin{figure}
\includegraphics[width=12cm]{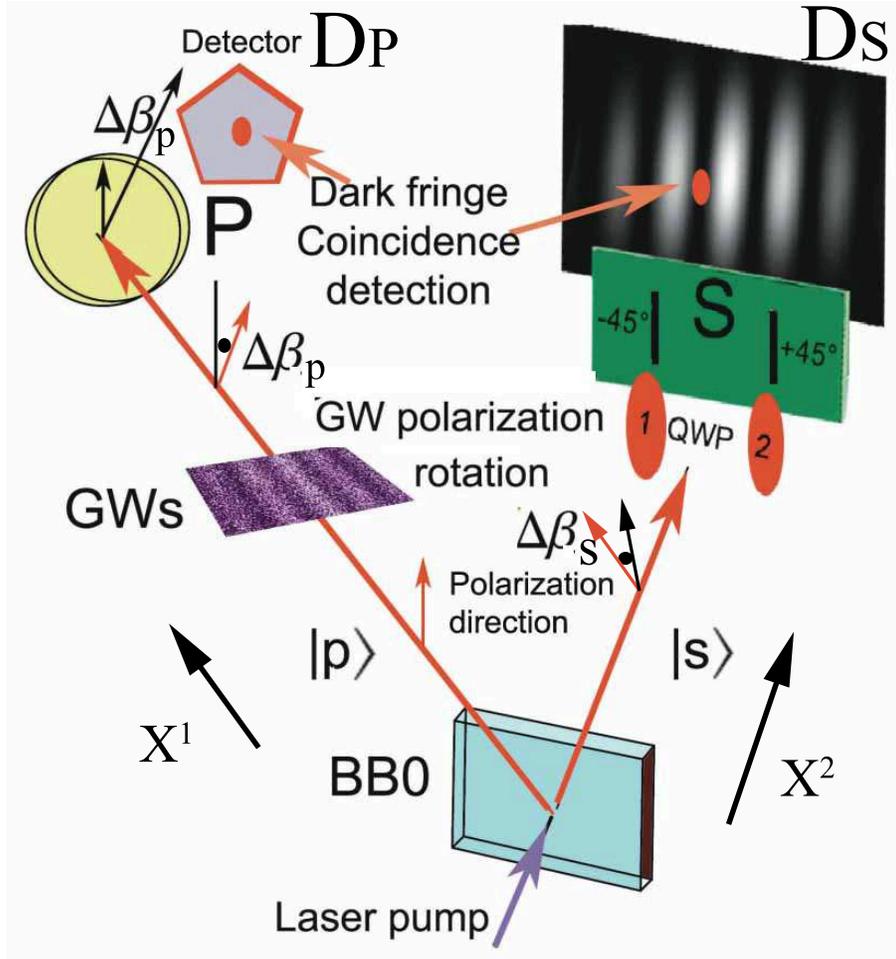}
\caption{Scheme of the GW detector: a laser beam on the BBO crystal creates pairs of entangled photons via a parametric downconversion process. One of the photons, $|s\rangle$ travels from the crystal to the double-slit S. In front of S are present two quarter-wave
plates QWP 1 and 2, with their fast axis mutually perpendicular. The other photon, $|p\rangle$, travels to the polarizer P located at a variable distance $l$ from the crystal. When the directions of the polarizer P and the fast axis of either QWP1 or QWP2 coincide, the interference pattern is restored, being canceled when we have only available the which-path information. If the setup interacts with a GW, the angles will not coincide any more, modifying the interference pattern.} 
\label{Fig.1}
\end{figure}

\section{Stochastic detector}

The experimental configuration discussed in point 2 above describes the situation in which the two QWPs mark the photon patterns after the double slit, leading to the destruction of the interference pattern. The polarizer P has been removed.
In the ideal case, when no GWs are present, the distributions of the coincidences counted by the detectors after the slit S and in the path p are always gaussian.
A parametric downconversion type II source produces photon pairs in the well-known linear superposition of ordinary and extraordinary polarizations. 

The presence of spacetime ripples or of continuous waves will introduce a rotation $\Delta \beta_s$ between the eigenvectors associated to the source (o and e) and the directions identified by the fast axes of the two QWPs. In this way, the ideal gaussian distribution obtained by counting the heralded photon coincidences is not gaussian any more.
In fact, the superposition of a gaussian distribution with a deterministic process like that generated by a plane GW, or the superposition with a different stochastic process, such as the stochastic GW background, is no longer gaussian \cite{tam08}.

The coincidence detection probability $P_1$ of eq.(\ref{e4}), in presence of gravitational waves then becomes
\begin{equation}
P_1\propto\frac{1}{2}(1+\sin\delta\cos(2\theta+2\Delta
\beta_s))
\label{e6}
\end{equation}
since $\Delta\beta\sim 0$, at linear order in the GW amplitude we have,
\begin{equation}
P_1\propto\frac{1}{2}(1+\sin\delta\cos(2\theta) -\sin \delta \sin (2 \theta) 2 \Delta \beta_s)
\label{eq11}
\end{equation}
since $\theta=\pm \frac{\pi}{4}$, eq (\ref{eq11}) becomes
\begin{equation}
P_1\propto\frac{1}{2}(1-2\Delta\beta_s\sin\delta). 
\label{e7}
\end{equation}
Here the photon $p$ is used only as herald for the photon $s$ that depicts a stochastic distribution as discussed. In addition to the classical GW detectors this setup favors the characterization of stochastic GW backgrounds. Anyway, the analogy of this last configuration with the already proposed entangled states GW detector is quite straightforward. Therefore for a deeper investigation and preliminary considerations about the presence of noise in the \textit{gedankenexperiment} we refer to the articles \cite{tam00,tam08}.

\begin{figure}
\includegraphics[width=14cm]{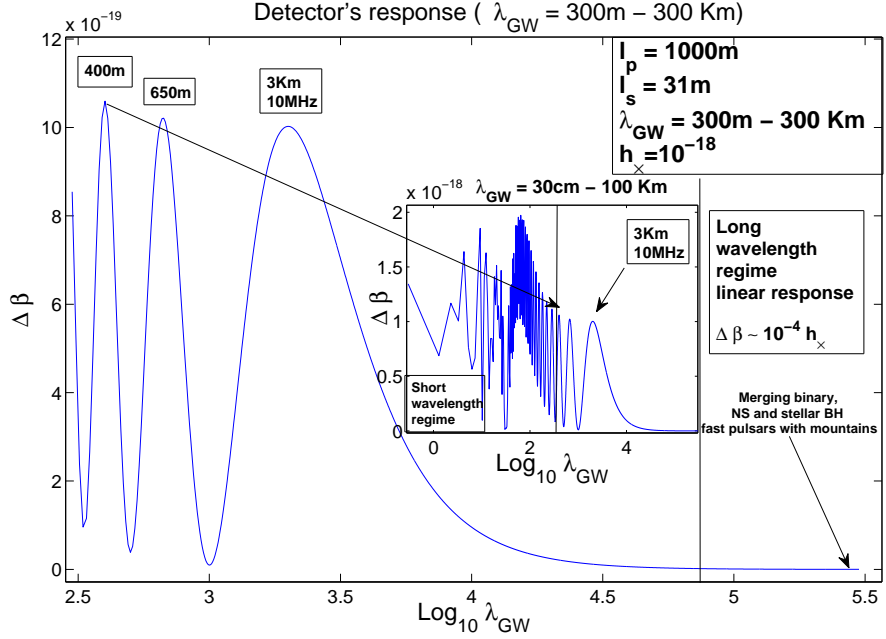}
\caption{Scheme of the GW detector: We now consider the high frequency spectrum of GWs in the GHz - KHz region. Here $l_p=1000m$ and $l_s=31m$. From eq. \ref{eqdbeta}  we obtain the response of the detector, $\Delta \beta$ as a function of a GW having $h_\times=10^{-18}$ in the spectrum of wavelengths $300m <\lambda_{GW}<300Km$. After the broader peak at 10MHz, the response stabilizes to an asymptotic regime that $\Delta \beta \sim 10^{-4} h_\times$, similar to the response obtained from the ground-based GW detectors without the limitation of the shot noise (see text). By varying the length of the path to the polarizer P ($l_p$) one can tune the detector at a specific GW wavelength also performing GW spectroscopy at ultra high frequencies. The three main peaks shift linearly with $l_p$.} 
\label{Fig.2}
\end{figure}

\section{Conclusions}

We propose to use heralded photons for the detection of gravitational waves.
The detection of one member of an entangled pair of photons can be used to herald the presence of the other photon. More in general, heralded photons offer the key advantage of certificating with high confidence, through the detection of their heralds, that a specific event has actually occurred.

Here we propose a \textit{gedankenexperiment} in which the use of heralded photons 
might discriminate a real event of a photon count in the dark fringes of the restored interference pattern of a Wallborn interferometer distorted by a gravitational wave from spurious random events.

Moreover, the use of heralded photons, as in the entangled photon case, permits to overcome the shot noise problem through a dependence of $1/N$ instead of $1/\sqrt N$, where $N$ is the number of the quanta in each multi-photon state. For a deeped discussion about the actual feasibility of measuring and building multi-entangled photon states in more efficient ways, see Refs. \cite{qi09,lu07,wie08}.

By applying Tamburini's \emph{\etal} concept for the detection of gravitational waves using quantum entanglement, we show that this proposal could be a direction worth being investigated in the near future especially for high frequency GWs.

We discuss the simplest case with $N=2$, more specifically the Walborn's quantum eraser experiment, in which the polarization properties of heralded photons pairs are used to destroy/restore the interference pattern after a double slit in where the which-way photon patterns are marked by the two quarter wave plates there present.

We point out that the choice of polarization entanglement is not the main key for the general use of heralded photons in GW detection. Heralded photons can be obtained also from entangled/correlated photon states different from polarization, such as energy-time or momentum-space, making possible their future implementation also in the classical interferometric detector setups.
The important result obtained from this \textit{gedankenexperiment} is to show that the deviation measured by using these particular quantum states is always the order the gravitational wave amplitude $h$, like in the classical GW detectors.
This shows that the proposal of improving the already existing detection techniques of GW interferometers (see e.g. VIRGO, LIGO, the Einstein Telescope ET) by adopting heralded photon coincidences instead of the photon counting obtained through classical laser sources is in principle feasible especially when enough bright sources will be available.

To conclude, EinsteinÕs theory of General Relativity reveals that gravitational energy is
not a usual form of ÒNewtonianÓ mechanical energy, since it is not possible
in the framework of this theory to obtain an expression for the energy of the
gravitational field satisfying simultaneously both the conditions that:
\begin{enumerate}
\item When added to other forms of energy is conserved and

\item The energy within a definite (three dimensional) region at a certain time
is independent of the coordinate system.
\end{enumerate}

Thus in general, as argued by Dirac, gravitational energy cannot be localized
\cite{dir96}. However, as is well known, for the case of
gravitational waves all moving only in one direction, gravitational energy
can be localized. In this particular case we thus have definite expressions
for the total energy and momentum, which are conserved and covariant. The
experimental setup we are proposing offers an innovative method in probing
the dual character of non-locality of gravitational radiation to the extent
that we can select the energy of the detected gravitational waves (by being
able to select the detection frequency), and we can evaluate how they depend
on the reference frames attached respectively to the polarizer and to the
double slit, by alternating between the stochastic and the non stochastic
detection methods. This last aspect could be relevant for the possibility to
use the proposed gravitational antenna to tackle the subject of preferred
frames in physics. Our experimental concept could also be pertinent to
investigate the relation between gravitational radiation non-locality in
General Relativity and the issue of non-locality in Quantum Mechanics since a
quantum Eraser uses entangled polarized photons.
We would also like to draw the attention of the reader to the natural
extrapolation of our experimental concept in the framework of a space
mission, which would consist, very roughly, of three spacecraft: one hosting
the entangled photon source, a second one hosting the polarizer, and a third
one hosting the double slit.

\acknowledgments
One of us, FT, also gratefully acknowledges the financial support from the 
CARIPARO Foundation inside the 2006 Program of Excellence.

\end{document}